# Thousands of *DebitCredit* Transactions-Per-Second: Easy and Inexpensive

Jim Gray, Microsoft Research
Charles Levine, Microsoft SQL Server

Jim Gray
*Microsoft Research*

# Thousands of *DebitCredit* Transactions-Per-Second: Easy and Inexpensive

Jim Gray, Microsoft Research
Charles Levine, Microsoft SQL Server

**Abstract:** *A $2k computer can execute about 8k transactions per second. This is 80x more than one of the largest US bank's 1970's traffic – it approximates the total US 1970's financial transaction volume. Very modest modern computers can easily solve yesterday's problems.*

## 1. A Thousand-Transactions-per-second was once difficult and expensive.

In 1973, Bank of America wanted to convert their paper-based branches, tellers, and demand-deposit (savings) accounts to an online system, letting tellers perform a customer's deposits and withdrawals. The corresponding transaction profile, called *DebitCredit,* evolved to become a standard measure of transaction processing [Serlin].

At the time, the system of ten thousand tellers needed to perform 100 transactions per second. The ten million account records were about 1GB and the 90-day general ledger was about 4GB. At the time, the server hardware for such a system cost more than ten million dollars; but, it was not until 1976 that a commercial database system was able to run 100 transactions per second [Gawlick].

A decade later, Tandem used a 34-CPU 86-disk SQL system costing ten million dollars to process 208-transactions per second. At the time, this was considered a breakthrough because relational systems had a reputation for poor performance [Tandem].

For much of the 1980's the database and transaction processing performance agenda was to achieve a thousand transactions per second. Part of that process defined the "one-transaction per second" unit. Informal definitions [Datamation], [1Ktps] and bench-marketing eventually led to the formation of the Transaction Processing Performance Council (www.tpc.org) which defined the TPC-A transaction profile largely in line with DebitCredit [Serlin]. By early 1990 several database systems had achieved the 1,000 tps milestone. By the late 1990's, clusters of 100 machines were delivering over 10,000 tpsA [Scalability]. Long before then, TPC-A was replaced by the more challenging TPC-C benchmark [TPC-C], [Levine].

TPC-C had a similar experience. The early systems delivered 1k tpmC at 2000$/tpmC, today systems are delivering about to 3M tpmC for about 5$/tpmC.

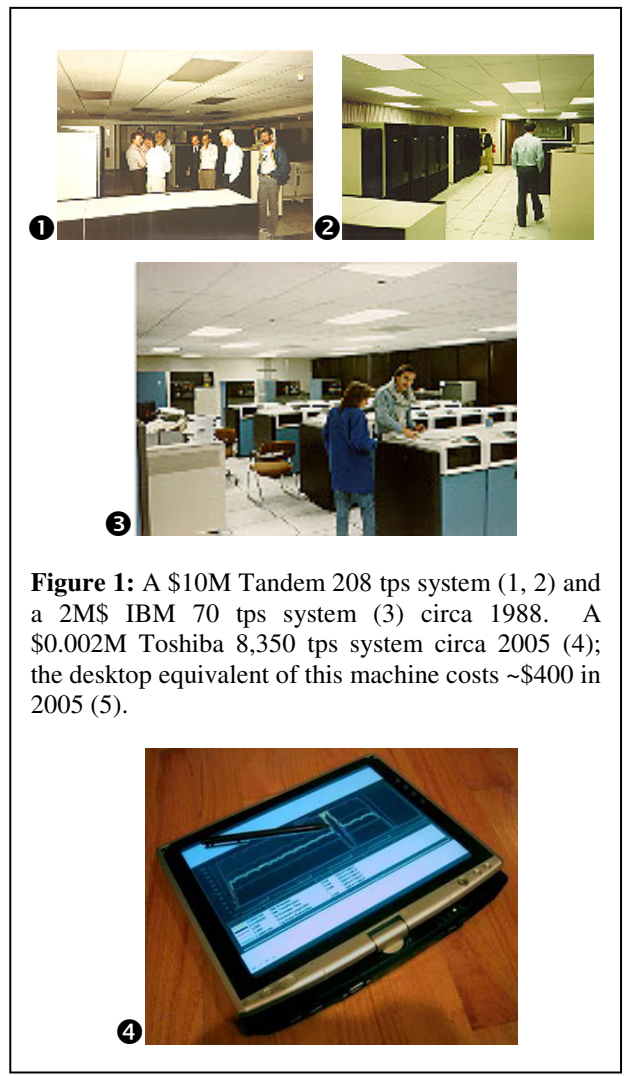

**Figure 1:** A $10M Tandem 208 tps system (1, 2) and a 2M$ IBM 70 tps system (3) circa 1988. A $0.002M Toshiba 8,350 tps system circa 2005 (4); the desktop equivalent of this machine costs ~$400 in 2005 (5).

## 2. DebitCredit on a PC?

We conjectured that a year 2003 PC could do ALL the 1970 U.S. banking transactions and store all the 1970 US bank accounts. Is that really true? In the early 1970s Bank of America did 10% of the U.S. banking transitions so the total would be 1,000 transactions per second and 100M accounts.

A Toshiba Protégé M200 TabletPC (see Figure 1.4) has a 1.6 GHz Centrino™ CPU with 2MB L2 cache, 1GB PC2700 RAM, a Hitachi 5K80 74GB 5400 RPM 2.5" ATA disk with 8MB write-enabled cache, and runs WindowsXP TabletPC 2005 SP2 with Microsoft SQLServer 2000 SP2. We wrote the following programs (see Appendix) for it:
(1) Create and populate a SQLServer DebitCredit database with 10M accounts.
(2) A DebitCredit database transaction (no message handling).
(3) An *N*-stream DebitCredit test harness.

These programs miss a few DebitCredit requirements: there is no message handling; the accounts all fit in RAM; and the log and database are on the same disk. Those issues are discussed in the next section, but here are the measurements.

It takes 365 seconds to create the database and log files and 173 seconds to populate the 10M accounts and 10,000 tellers. So, the benchmark takes ten minutes to set up. Randomly warming up the cache (getting all 250MB of the accounts into memory) from a cold start takes about 4 minutes (Figure 2A) – it takes less than 15 seconds if done as a sequential scan. The 10M account records (of about 25 bytes each) occupy about 250MB on disk and RAM. Overall, the database and database server have a working set of about 325 MB. At equilibrium the account, branch, and teller records are in main memory - there are no more reads of the database from disk. SQLServer is writing the log and the history table. Figure 2.A shows the dynamics of ramp up and of adding more request streams (threads). As threads are added one begins to see group-commit and CPU savings from shared commit processing.

A single-threaded DebitCredit run uses about 70% of the processor, averages 2,250 transactions per second, and has a peak rate of 2,635 tps (Figure 2.B). In the single-threaded case each transaction is flushing a partially filled 512-byte log page. That causes 30% CPU wait time. Checkpoints are configured to limit recovery time to 10 minutes. Notice that checkpoint causes throughput to drop.

A multi-threaded run averages over 8,346 transactions per second and is CPU bound (actually memory latency bound). The 8,923 tps peak rate suggests each transaction costs ~112 CPU microseconds (~191k clock ticks). CpuMon indicates 1.9 CPI and so ~100 k instructions per transaction The graph shows approximately 3 transactions per log force, about 500 log bytes per transaction, and the disk controller saturating at 3,500 requests per second. As threads are added one begins to see group-commit and CPU savings from shared commit processing.

**A.** Throughput ramps-up as accounts become memory resident and stabilizes at 2,500 tps and 70% CPU utilization. At that point log writes are the only IO activity (checkpoints are disabled for this run). A 2[nd] thread gives 5,000 tps and doubles log writes. A 3[rd] thread gives 6,500 tps and 1.5 transactions per log write (some group commit.) By the 7[th] thread, peak is 8,843 tps, CPU is saturated, and there are 2.5 transactions per log write.

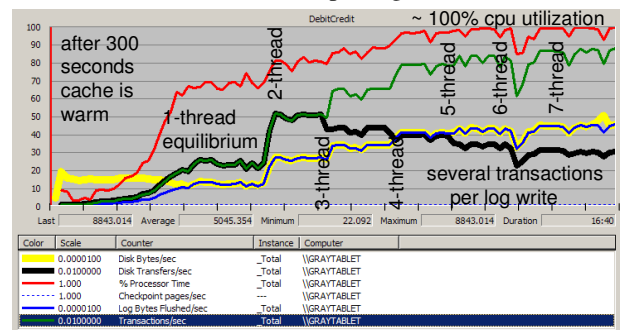

**B.** 1-thread, 2M transactions, 15 minute warm run.

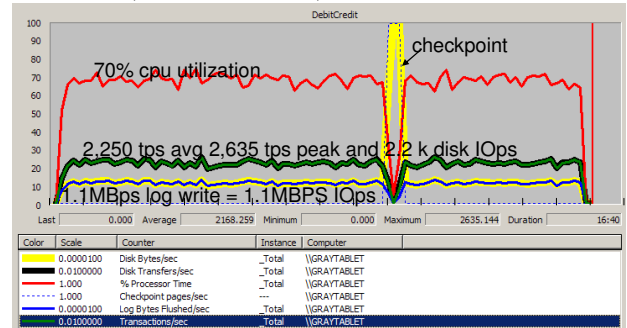

**C.** 8-thread warm run with checkpoint.

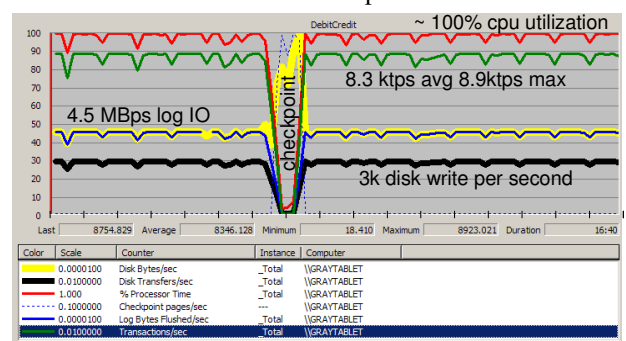

**Figure 2:** Performance monitor graphs. Samples every 10 seconds: red: CPU, green: tps, black: IO/s, yellow: bytes/sec, blue: log flushes/s.

## 3. Caveat: Why these tps results are bogus

Ten million customer accounts and 90 days of history occupy 5 GB and fit comfortably on the computer's 74 GB disk – but that is a 100tps system. According to the DebitCredit (and TPC-A, TPC-B) scaling rules [Serlin], 8,300 tps implies a bank of 830 million customers. That is 420GB of disk space and at least 80GB of RAM. So, this is not a scaled implementation. But, 400GB disks exist, RAM prices are coming down, and 64bit addressing has arrived (even for laptops). Indeed, the system shown in Figure 1.4 is equivalent to a $400 desktop system which could be expanded to enough disk and RAM to hold a fully scaled 1K tps system. But, an 80 GB server is likely a quad Opteron that can do many more than 8k tps. So, to repeat the earlier comments, the system will be disk and memory bound, not CPU bound.

The TPC-A rules specified that each transaction has a 100 byte input message and a 100 byte output message. Adding that logic would soak up some CPU and make the TabletPC a few ktps system.

TPCC-A and DebitCredit require *durability*. The laptop battery and Windows hibernation might pass the TPC auditors durability test. But the auditor certainly would want the log to be on a separate and duplexed device so that it would mask single media failures; so that if the database disk failed, or if one of the log disks failed, the customer could recover all committed transactions from the surviving disks and from the offline database backup files. One could easily add those disks but…

TPC-A rules also specify that 90% of the transactions should have less than 2 second response time. That rule was designed to prevent the kinds of performance dips shown in Figure 2. 90% of the transactions in these runs did indeed have response time less than 2 seconds – but the 20 and 30 second "blackouts" during the 900 second runs in Figure 2 are troublesome. In a properly scaled and configured TPC-A run, we would have 830,000 terminals each submitting transactions every 100 seconds. The 20-30 second blackout would cause ~250,000 transactions to exceed the 2 second response time and that would have ripple effects. But 250k transactions is less than 4% of the 7M transactions run in the 15 minute test window – so if the ripple effects were minor, the blackout would just be embarrassing, not disqualifying. One of the deficiencies of TPC-B and of this work is that it doesn't have terminals and so it fails to properly model these effects.

## 4. Summary and Observations

The next article in this series, scheduled for April fools day 2025, will show that a $1 wrist watch can run the world economy as of 1990. Since cell-phones are already at a gigabyte of storage and approach a GHz processor, such an article may be possible – we hope we are around to write it.

Moore's law forces give approximately a 100x price/performance improvement each decade. This progress is a combination of hardware improvements (processors, memory, disks, and networks) from software improvements (algorithms,) and from changes in business models (commoditization). Figure 3 shows this trend for the TPC benchmarks which have a wealth of audited price-performance results. The graphs show a trend line that has a 100x improvement per decade. That translates to 58% per year tps/$ improvement and consequent 37%/year price reduction in $/tps over 15 years.

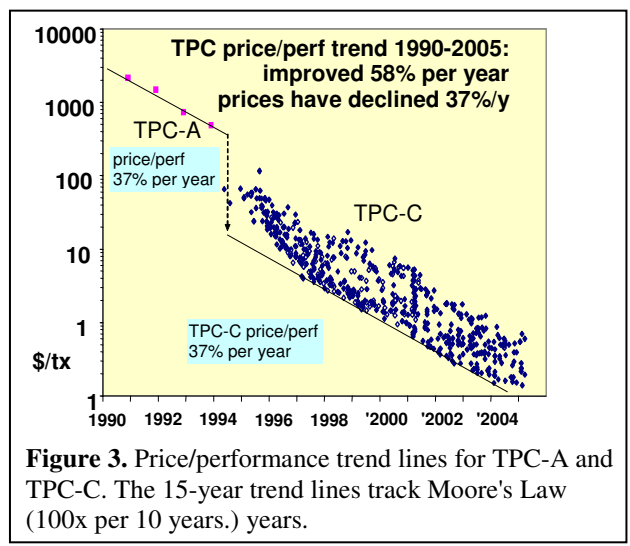

**Figure 3.** Price/performance trend lines for TPC-A and TPC-C. The 15-year trend lines track Moore's Law (100x per 10 years.) years.

The first TPC-A/B systems were in the $100k/tpsA range. By 1995 they were approximately $500/tpsA. The measurements here seem to be below 1$/tpsA, even after you factor in the correct scaling rules and the costs of delivering messages (excluding the cost of the terminals mandated by the TPC-A rules).

As a result of this trend, the impossible 1970 task became a $10M 1990 task and the $10K task of 2005. The old performance problems are easy to solve with today's computers.

The ability to do 8ktps on a laptop demonstrates that you *can* use relational systems and simple algorithms if your problem involves a few thousand transactions per second per CPU and if your data fits in RAM.

The results also indicate that CPU is not the problem – if you can feed memory to the CPU, it delivers impressive transaction rates. Most processors are stalled waiting for RAM, network, or disk – even this laptop is memory limited – the CPI of 1.9 suggests that it is waiting for memory about half the time.

The results also show that it is important to pay attention to massive main memory and checkpoint performance. Checkpoint IO should be spread across the whole checkpoint interval, rather than rushing to do it all at once. As Figure 2 shows, SQLServer 2000 keeps the IO queue 100 requests deep until the checkpoint completes – this starves the other tasks – most systems have a dedicated log disk and can service a 100-deep IO queue, so this is not a problem. The next version of SQLServer fixes this problem by reducing the outstanding checkpoint IO if the IO response time increases dramatically.

The main point, for anyone involved in the benchmark wars of the 1980s and 1990s, is to marvel at the power of modern systems. They solve the old performance problems, leaving us free to focus on the many new performance problems. If we IT folks had the luxury of generals who fight the previous war, life would be boring. Although we do not have the DebitCredit problem anymore, it is nonetheless marvelous that we can solve it so easily.

## *Appendix 1: Debit Credit Sample Code*

Create the database and define the database schema

```
-------------------------------------------------------------
-- create the database files and the database metadata. (365 seconds)
set nocount on
create database theBank
    ON    ( name = data, FILENAME = 'c:\TheBank\Data.mdf', SIZE = 1GB)
    LOG ON ( name = log,  FILENAME = 'c:\TheBank\log.ldf',  SIZE = 5GB)
alter database theBank set recovery simple
exec sp_configure 'recovery interval', '10'  -- recovery takes at most 10 minutes
exec sp_configure 'max server memory', '325' -- limit working set to 325 MB
reconfigure with override
go
-------------------------------------------------------------
-- after 130 seconds, connect to it.
use theBank
go
-------------------------------------------------------------
-- create the branch, teller, account, history tables.
-- they are not padded to the 100-bytes of tpcA
--   (so that they will fit in 1/2GB of ram on my tabletPC.)
-- 1000 branches is 10M customers and about 300 MB.
create table Branch(   branchID    int   not null primary key,
                       balance     float not null )
create table Teller(   branchID    int   not null
                               foreign key references Branch(branchID),
                       tellerID    int   not null primary key,
                       till        float not null )
create table Account(  branchID    int   not null
                               foreign key references Branch(branchID),
                       accountID   int   not null primary key,
                       balance     float not null )
create table History(  timestamp datetime not null default getdate(),
                       branchID    int   not null,
                       tellerID    int   not null,
                       accountID   int   not null,
                       amount      float not null )
go
```

Next it is time to populate the database with the 10M accounts.

```
-- ==========================================
-- Fill bank database: each branch has 10 tellers, 10,000 accounts.
-- account numbers have branch encoded in "millions" part of account ID.
-- ==========================================
-- create a store procedure to fill the Bank
create procedure spFillBank @branches int as
   begin
   begin transaction
   -- First empty the tables so we are starting fresh
   delete Account; delete History; delete Teller; delete Branch
   commit transaction

   declare @branchID int, @tellerID int, @accountID int
   declare @tellersPerBranch int, @accountsPerBranch int, @BranchRadix int
   set @tellersPerBranch = 10        -- 10 tellers per branch
   set @accountsPerBranch = 10000    -- 10 thousand accounts/branch
   set @BranchRadix     = 1000000 -- 1m is radix for branch in acct/teller ID
 -------------------------------------------------------------
   -- for each branch, start a transaction and make its accounts and tellers
   set @branchID = 0
   while (@branchID < @branches)
      begin
      begin transaction
      -- the branch record
      insert Branch values (@branchID, 0.0)
      set @tellerID = 0
      -- add branch's 10 teller records (teller ids have branch id at radix 1m)
      while (@tellerID < @tellersPerBranch)
         begin -- teller ID = | BranchID  | TellerSeqenceNumber |
         insert Teller values(@branchID,@branchID*@BranchRadix+@tellerID, 0.0)
         set @tellerID = @tellerID + 1
         end
      set @accountID = 0
      -- add 10k account records (account ids have branch id at radix 1m)
      while (@accountID < @accountsPerBranch )
         begin  -- account ID = | BranchID  | AccountSeqenceNumber |
        insert Account values(@branchID,@branchID*@BranchRadix+@accountID,0.0)
         set @accountID = @accountID + 1
         end
      set @branchID = @branchID + 1
      commit transaction
      end
   end
go
```

```sql
---------------------------------------------------------------
-- The classic database part of TPC-A (and DebitCredit)
-- This is a single DebitCredit database transaction.
create procedure spDebitCredit @tellerID int, @accountID int, @amount float as
    begin
    declare @newBalance    float,
            @branchID      int,
            @BranchRadix   int
    set @BranchRadix=1000000 -- 1m is branch radix in account/teller ID
    set @branchID = @tellerID / @BranchRadix
    begin transaction
    update Teller set till = till + @amount
                        where tellerID = @tellerID
    update Account set @newBalance = balance = balance + @amount
                        where accountID = @accountID
    insert History (branchID, tellerID, accountID, amount)
          values ( @branchID, @tellerID, @accountID, @amount)
    update branch set balance = balance + @amount
                        where branchID = @branchID
    commit transaction
    end
 go
```

```sql
---------------------------------------------------------------
-- Run N DebitCredit Transactions (picking random accounts).
create procedure spRunDebitCredit @transactions bigint as
   begin
   ------------------------------------------------------
   -- Global varibles and constants.
   declare  @branchID int, @tellerID int, @accountID int, @amount float
   declare  @branches int
   declare  @tellersPerBranch int,
            @accountsPerBranch int,
            @BranchRadix int
   select @Branches= count(*) from Branch with(nolock); -- 1,000 branches
   set @tellersPerBranch   = 10      -- 10 tellers per branch
   set @accountsPerBranch  = 10000   -- 10 thousand accounts/branch
   set @BranchRadix    = 1000000 -- 1m is branch radix in acct/telr ID

   --------------------------------------------------------------
   -- do @transactions transactions picking a random account.
   -- Pick a random teller at a random branch (15%/85%) remote/local
   while @transactions > 0
      begin
      set @branchID = rand()*@branches -- random branch with rand teller
      set @tellerID = @branchID*@BranchRadix + rand()*@tellersPerBranch
      if (rand() >= .15)         -- 85% account is local to branch
         set @accountID = @branchID*@BranchRadix
                                 + (rand()*@accountsPerBranch)
      else              -- 15% non local accounts
         set @accountID = floor((rand()*@branches))*@BranchRadix
                                 + (rand()*@accountsPerBranch)
      set @amount = rand()*1000 – 500  -- deposit between –500$ nd 500$
      -- parameters computed, now do DebitCredit.
      exec spDebitCredit @tellerID, @accountID, @amount -- do it
      set @transactions = @transactions – 1  -- decrement tran count
      end             -- bottom of transaction loop
   end
   return
go
```

```sql
---------------------------------------------------------------
-- do the single-threaded benchmark and time 2m transactions
---------------------------------------------------------------
--      Create a database of 1000 branches
exec spFillBank 1000
checkpoint
-- takes 173 seconds and produces a 250 MB database on SQL2K
go
------------------------------------------------------
-- now do the benchmark (note the db creating primed the cache).
declare @transactions bigint
---------------------------------------------------------------
-- prime the cache, so a warm start.
select count(*)from Account; select count(*)from Teller;
truncate table history
checkpoint
go

---------------------------------------------------------------
-- initial clocks: used as basis for timing
declare @clock datetime, @cpu bigint, @physical_io bigint, @elapsed bigint
select    @clock       = getdate(),         -- read current time, and IO
          @cpu         = @@CPU_BUSY ,
          @physical_io = @@TOTAL_READ + @@TOTAL_WRITE

---------------------------------------------------------------
-- run the test for a 2m transactions
set @transactions = 2000000
exec spRunDebitCredit @transactions

------------------------------------------------------------
-- gather performance at the end
Select @elapsed  = datediff(ms, @clock, getdate()),
       @cpu         = (@@CPU_BUSY - @cpu) * @@TIMETICKS,-- scale ticks/ microsecond
    @physical_io = (@@TOTAL_READ + @@TOTAL_WRITE) – @physical_io
-- treat wraparound of counters as a zero value
if @cpu < 0 set @cpu = 0; if @physical_io < 0 set @physical_io = 0
-- printout
print ' ran ' + cast (@transactions as varchar(30)) + ' transactions '
      + ' cpu: '         + str(@cpu/1000000.0, 8,0)
      + ' sec, elapsed: '   + str(@elapsed/1000.0,8,0)
      + ' sec, physical_io: ' + str(@physical_io,8,0)
      + ' tps: '         + str(@transactions/(@elapsed/1000.0), 8,0)
```

***Appendix 2:*** `ParallelBatch.bat` *script to run N DebitCredit Threads in Parallel*

The following is the script used to evaluate the "staged" increase in threads in Figure 2A. Figure 2.C was generated by removing the sleep commands. The commands launch a SQL command to run the `spRunDebitCredit` stored procedure with a parameter of 2 million on the `TheBank` database with windows security (`-E` requests windows security).

```
start isql -E -d TheBank -Q "EXIT(exec spRunDebitCredit 2000000)" > c:\sqllog01.txt
sleep 360
start isql -E -d TheBank -Q "EXIT(exec spRunDebitCredit 2000000)" > c:\sqllog02.txt
sleep 100
start isql -E -d TheBank -Q "EXIT(exec spRunDebitCredit 2000000)" > c:\sqllog03.txt
sleep 100
start isql -E -d TheBank -Q "EXIT(exec spRunDebitCredit 2000000)" > c:\sqllog04.txt
sleep 100
start isql -E -d TheBank -Q "EXIT(exec spRunDebitCredit 2000000)" > c:\sqllog05.txt
sleep 100
start isql -E -d TheBank -Q "EXIT(exec spRunDebitCredit 2000000)" > c:\sqllog06.txt
sleep 100
start isql -E -d TheBank -Q "EXIT(exec spRunDebitCredit 2000000)" > c:\sqllog07.txt
sleep 100
start isql -E -d TheBank -Q "EXIT(exec spRunDebitCredit 2000000)" > c:\sqllog08.txt
```